\documentclass[aps, twocolumn, prl, showpacs,preprintnumbers,amsmath,amssymb,superscriptaddress]{revtex4-1}
\usepackage{amsmath,amsfonts,amssymb,graphics,graphicx,epsfig,color,times,indentfirst,layout,hyperref,bm}
\usepackage{ulem}
\hypersetup{linktocpage,,colorlinks=true}

\def\k{{\bf k}}
\def\C{\mathcal{C}}
\def\H{\mathcal{H}}
\def\L{\mathcal{L}}
\def\S{\mathcal{S}}

\begin{document}

\title{Weyl-link semimetals}

\author{Po-Yao Chang}
\email{pychang@physics.rutgers.edu}
\affiliation{Center for Materials Theory, Rutgers University,
Piscataway, New Jersey, 08854, USA }

\author{Chuck-Hou Yee}
\email{chuckyee@physics.rutgers.edu}
\affiliation{Center for Materials Theory, Rutgers University,
Piscataway, New Jersey, 08854, USA }

\begin{abstract}
  A family of topological semimetallic phases where two-fold degenerate gapless
  points form linked rings is introduced. We refer to this phase as Weyl-link
  semimetals. A concrete two-band model with two linked nodal lines is
  constructed. We demonstrate that the Chern-Simons 3-form depends on the
  linking number of rings in a generic two-band model. In addition, we show the emergence of zero-energy
  modes in the Landau level spectrum can reveal the location of nodal lines,
  providing a method of probing their linking number.
\end{abstract}

\maketitle

\section{Introduction}

The application of topology to condensed matter physics has produced rich
insights into the behavior of an entire class of materials. The integer quantum
Hall state, as well as topological insulators and
superconductors~\cite{konig07, hsiehNature08, hasanKaneReview10,
  qiZhangReview10, HasanMooreReview11, ryuNJP10}, are characterized by the
topology of their ground state wave function. These topological structures give
rise to protected gapless boundary modes and quantized electromagnetic and
gravitomagnetoelectric responses~\cite{zhangPRB08,Nomura}. In addition to fully
gapped topological phases, semimetals and nodal superconductors can also
exhibit interesting topological properties. Of particular interest are Weyl
semimetals~\cite{WengPRX, LiuPRB, 2015arXiv150203807X}, where the band touching
points in the bulk behave as monopoles in momentum space. These monopoles are
sources and drains of Berry flux, which leads to anomalous electromagnetic
transport and the emergence of surface Fermi arcs. In addition to Weyl
semimetals, the set of gapless points in the bandstructure can also form
one-dimensional nodal lines and rings~\cite{Burkov2011, Matsuura2013, Fang2015,
  Zhao2017}. These nodal-ring semimetals and superconductors also exhibit
robust drumhead surface states~\cite{Matsuura2013, Bian2016}. Many material
candidates have been proposed and some have been experimentally
confirmed~\cite{Yu2015, Kim2015, Bian2016_1, Xu2017, Lu2016, Hu2016,
  Schoop2016, Singha2016, Chen2017_1, Wang2016, Neupane2016}.

In general, the complexity of gapless phases is richer than gapped phases in
the following sense: gapped phases are like a featureless vacuum while the
point and line nodes in gapless phases behave as defects in momentum space
carrying topological charge. As mentioned above, point nodes behave like
monopoles of Berry flux, while line nodes are akin to flux tubes (or
solenoids). The interplay between these momentum defects often leads to
observable effects. For example, nodal points and lines can coexist in the
momentum space~\cite{Goswami2015}. Nodal lines can intersect to form states
termed nodal-chains~\cite{Bzduvsek2016, Yu2017}. Finally, the line defects can
share termination points which can be seen as the momentum space equivalent of
the real space nexus previously discussed in helium-3~\cite{Hyart2016,
  Heikkila2015, Gao2017}.

We extend the family of gapless phases by constructing a minimal two-band model
containing two linked nodal rings, with the linking number controlled by an
integer $n$. We refer to this phase as a Weyl-link semimetal (WLSM). Similar to
a nodal-ring semimetal, the non-vanishing Berry phase around the nodal rings in
a WLSM leads to drumhead surface states. We show the Chern-Simons 3-form can
serve as a topological invariant of the linking number of nodal rings in the
two-band model. In the context of three-dimensional (3D) topological insulators,
the $\theta$-angle in the axion Lagrangian is related to the Chern-Simons
3-form, where the axion Lagrangian~\cite{zhangPRB08} is,
\begin{equation}
  \S_{\theta}^{\rm EM} = \frac{\theta e^2}{2 \pi h c}
  \int \! dt\, d^3x\, {\bf E \cdot B}.
  \label{Eq:axion}
\end{equation}
Since the axion Lagrangian can also be present in semimetallic phases such as
Weyl semimetals~\cite{Chen2013, Son2013, Landsteiner2014}, the nonvanishing
Chern-Simons 3-form in the WLSM potentially may affect magnetoelectric
transport. In addition, we analyze the Landau level spectrum in the WLSM and
observe the emergence of zero-energy modes. The zero-energy modes in the Landau
level spectrum reflect the locations of the nodal rings. Thus, the Landau level
spectrum provides a method of probing whether the nodal lines are linked or
not, similar to the use of quantum oscillation measurements for mapping out the
topology of Fermi surfaces in metals.

\section{Two-band models and surface states}

\begin{figure}
  \includegraphics[width=\columnwidth] {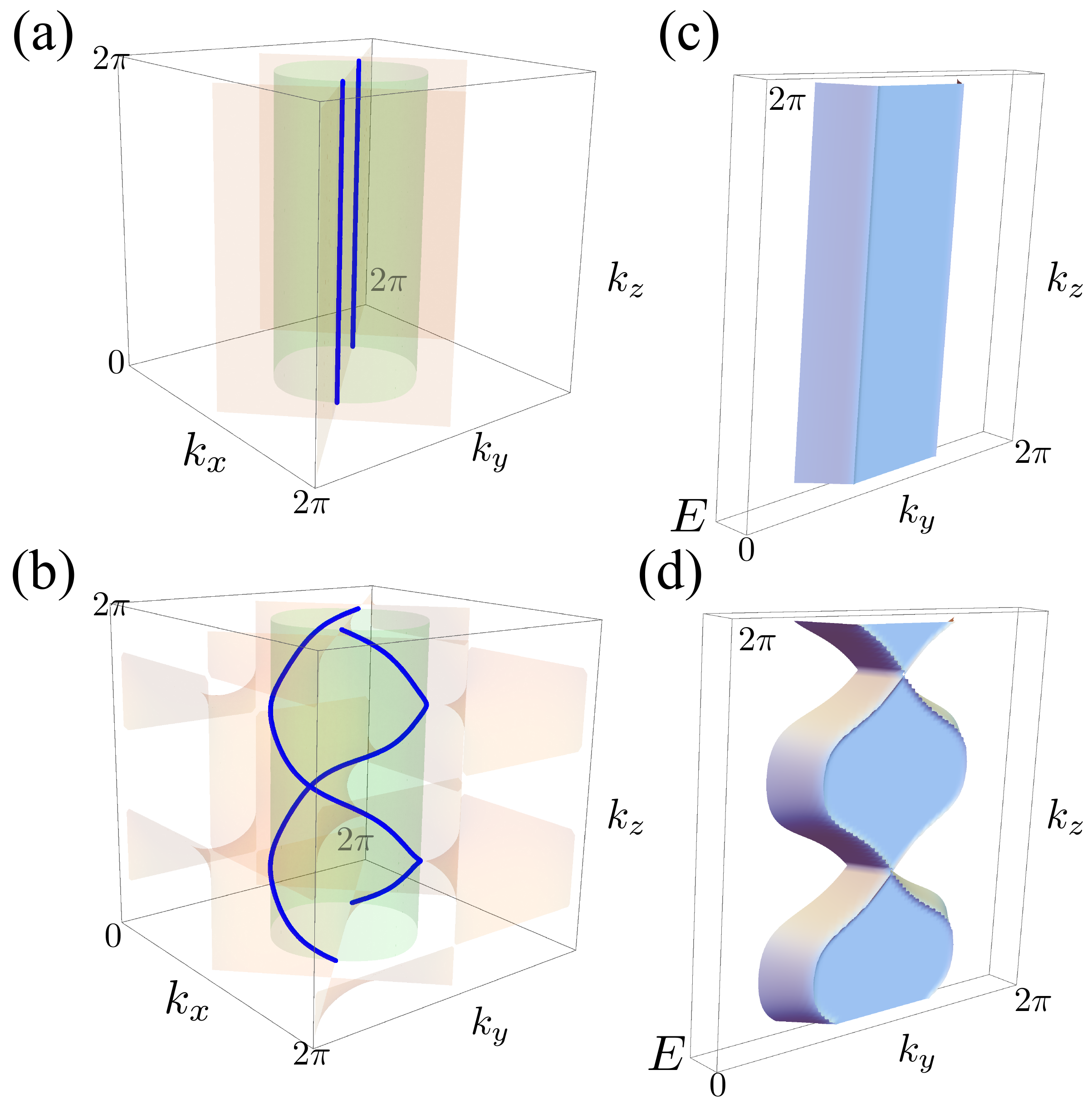}
  \caption{(a) Weyl-link semimetal (WLSM) with linking number $n = 0$. The red
    and green surfaces are the set of points where $f(\k) = 0$ and $g(\k) = 0$
    respectively. The nodal lines (blue) are the intersections between these
    two surfaces. (b) Energy spectrum of the $(100)$ surface states: the flat
    surface bands are bounded by the projected bulk nodal lines. (c) The WLSM
    with $n = 1$, an example of two linked nodal lines, and (d) its
    corresponding surface states. The parameter $t = 1$ for both model in
    Eq.~(\ref{Eq:f}).}
  \label{F1}
\end{figure}

We construct a two-band model containing two Weyl rings which link an arbitrary
number of times by considering the hamiltonian
\begin{equation}
  \H(\k) = f(\k) \sigma_x + g(\k) \sigma_y,
  \label{Ham}
\end{equation}
where $\sigma_x$ and $\sigma_y$ are Pauli matrices. This is a valid hamiltonian
for any periodic $f(\k) $ and $g(\k)$ in the Brillouin zone (BZ). The
energy spectrum is $E_{\pm}(\k)= \pm \sqrt{f^2(\k)+g^2(\k)}$ and the
corresponding eigenstates are
\begin{equation}
  |u_{\pm}(\k) \rangle = \frac{1}{\sqrt{2 E^2_{\pm}(\k)}}
  \begin{pmatrix}
    \pm f(\k)\mp i g(\k) \\
    |E_{\pm}(\k)|
  \end{pmatrix}.
\end{equation}
The two bands touch whenever both $f$ and $g$ vanish. For a general 3D system,
the zeros of these functions form 2D surfaces, and their intersection will form
1D nodal lines in the BZ. For a concrete realization of a WLSM, we choose the
following:
\begin{align}
  f(\k) &= t(1+\cos k_x +\cos k_y), \notag\\
  g(\k) &= t [\cos (n k_z) + \sin (n k_z) ] \sin k_x  \notag\\
  &+ t [\cos (n k_z) - \sin (n k_z) ] \sin k_y, 
  \label{Eq:f}
\end{align}
The surfaces of zeros of $f$ and $g$ intersect and form the desired Weyl links,
as shown in Fig.~\ref{F1}. The integer $n$ determines the (signed) linking
number, and negative $n$ corresponding to the opposite helicity in the $k_z$
direction. For $n = \pm 1$, Weyl rings link once while winding around the BZ,
and is topologically equivalent to the Hopf link.

\begin{figure}
  \includegraphics[width=\columnwidth] {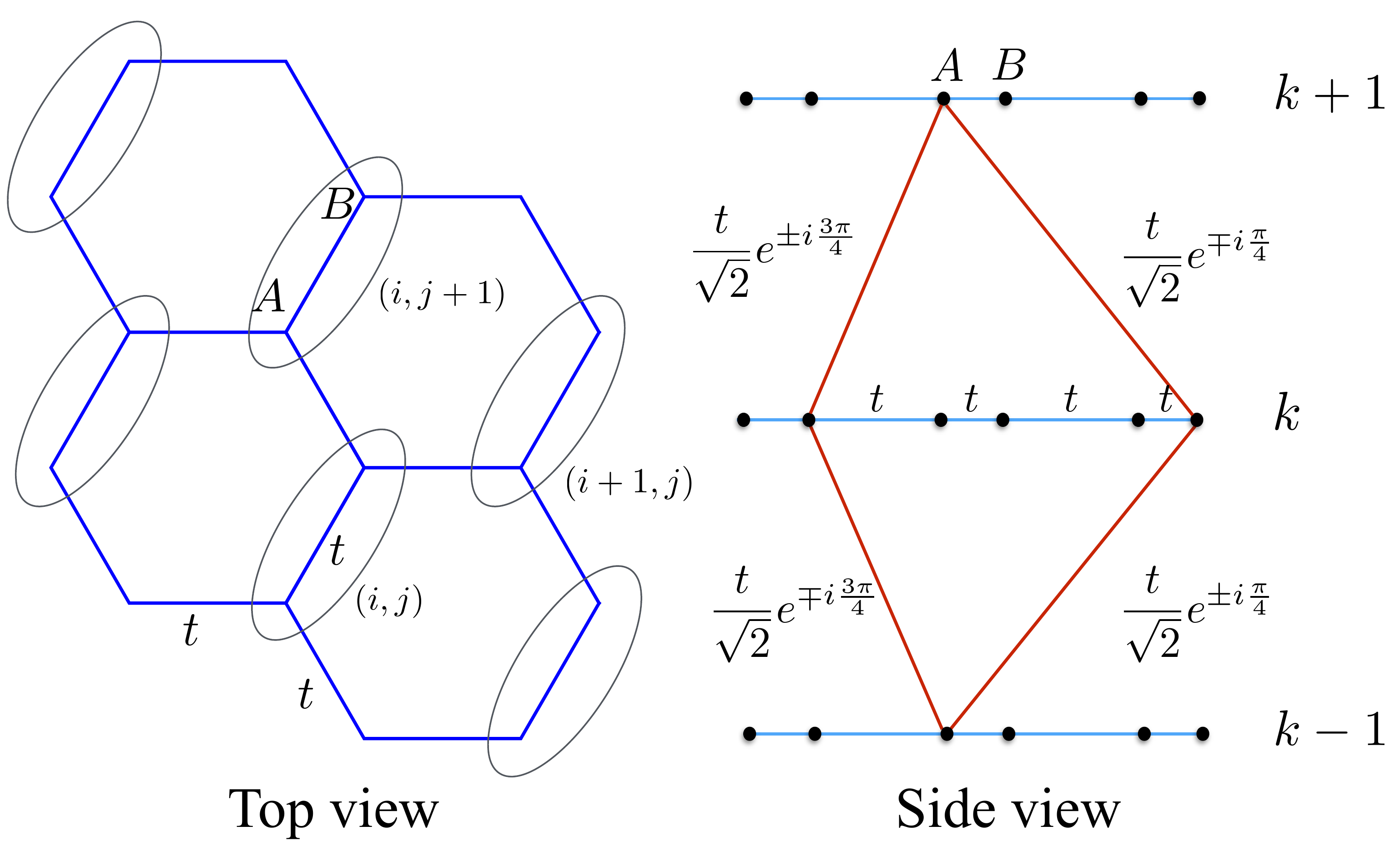}
  \caption{Tight-binding model for $n=1$ case. The top view indicates a graphene
    like structure with $A$ and $B$ sites per unit cell and the hopping
    amplitude $t$. The side view shows the inter-layer hopping with additional
    phases $e^{\pm i \pi/4}$ and $e^{\pm i 3\pi/4}$ as described in
    Eq.~(\ref{Eq:TB}).}
  \label{F0}
\end{figure}

One may worry about the mass term $m(\k)\sigma_z$ and shift $h(\k) \cdot
\mathbb{I}_{2 \times 2}$ present in generic 2-band models. The shift $h$ is
difficult to remove via symmetry constraints, but its presence does not gap out
the nodal lines, rather only shifting them in energy. If $h$ is small relative
to the other terms, the nodal lines will broaden to become small Fermi pockets
as the nodal line shifts above and below the Fermi level. The mass term $m(\k)
\sigma_z$ does gap out the nodal lines, and can be eliminated via symmetry
constraints. In the presence of chiral (sublattice) symmetry, the hamiltonian
must satisfy $\S^{-1} \H(\k) \S = -\H(\k)$ with $\S = \sigma_3$, thus
forbidding a mass term. The hamiltonian can also be rotated to form
$\tilde{\H}(\k) = f(\k) \sigma_x + g(\k) \sigma_z$. In this basis,
inversion symmetry combined with time-reversal symmetry ($\mathcal{PT}$) forces
the Hamiltonian $\tilde{\H}(\k)$ be real, forbidding the mass term $m\sigma_y$
and protecting the nodal lines~\cite{Burkov2011, Fang2015, Zhao2017}.

In principle, there is no symmetry requirement that forces the nodal lines to
be linked. However, the physical realization of the model proposed in
Eq.~(\ref{Eq:f}) is related to the Weyl points in spinless graphene. Let us
suppose we monitor the Weyl points in spinless graphene in a rotating frame
with the rotation axis perpendicular to the graphene and the rotation angle
being $n k_z$. The momenta $(k_x, k_y)$ will be transformed as $(\cos (n k_z)
k_x-\sin (n k_z) k_y, \sin (n k_z)k_x+\cos(n k_z) k_y)$. In this rotating
frame, two Weyl points will form two linked nodal lines along the $k_z$
direction.

The idea of rotating frame in graphene leads to a concrete tight-banding model
in the real space. In the case that the linking parameter vanishes $n=0$, the
tight-binding model of Eq. (\ref{Eq:f}) describes the layered graphene with
vanishing inter-layer hopping. In the case that the linking parameter $n=1$,
the tight-binding model of Eq. (\ref{Eq:f}) in the real space can be written as
\begin{align}
  H = &t \sum_{r} c^\dagger_{r,A} c_{r,B}+c^\dagger_{r+\hat{x},A} c_{i,B}+c^\dagger_{r+\hat{y},A} c_{r,B} \notag\\
  &+ \frac{1}{\sqrt{2}} (e^{i 3 \pi/4} c^\dagger_{r+\hat{x}+\hat{z},A} c_{r,B}+ e^{- i 3 \pi/4} c^\dagger_{r+\hat{x}-\hat{z},A} c_{r,B} \notag\\
  &+ e^{- i\pi/4} c^\dagger_{r-\hat{x}+\hat{z};A} c_{r,B}+ e^{i\pi/4} c^\dagger_{r-\hat{x}-\hat{z},A} c_{r,B} \notag\\
  &+ e^{-i3 \pi/4} c^\dagger_{r+\hat{y}+\hat{z},A} c_{r,B}+ e^{i3 \pi/4} c^\dagger_{r+\hat{y}-\hat{z},A} c_{r,B} \notag\\
  &+e^{i \pi/4} c^\dagger_{r-\hat{y}+\hat{z};A} c_{r,B}+ e^{-i \pi/4} c^\dagger_{r-\hat{y}-\hat{z},A} c_{r,B} ) \notag\\
  &+\text{h.c.}
  \label{Eq:TB}
\end{align}
The first three terms describe the hopping between $A$ and $B$ sites within the
same layer as the tight-binding model for a single graphene sheet. The other
terms describe the inter-layer hopping between $A$ and $B$ sites as shown in
Fig. \ref{F0}.

The nodal lines behave like vortex tubes in momentum space. The vorticity is
the topological charge of the line, and is given by the Berry phase divided by
$2\pi$,
\begin{equation}
  \nu_{\alpha}= \frac{1}{2 \pi}\oint_{\C_{\alpha}} a,
  \label{Eq:vorticity}
\end{equation}
where the Berry connection is
\begin{equation}
  \begin{split}
    a &= i\langle u_-(\k)| \partial_{k_i} u_-(\k) \rangle d k_i \\
    &= \frac{g\partial_{k_i}f - f\partial_{k_i}g}{2 (f^2 +g^2)}d k_i,
  \end{split}
\end{equation}
and $\C_{\alpha}$ is a loop encircling $\alpha$-th nodal line. Since the Berry
phase is defined modulo $2\pi$, the topological charge of a single nodal line
is either $+1/2$ or $-1/2$, depending on the relative orientation of the
probing loop $\C_\alpha$ versus the vorticity direction of the line
$\L_\alpha$.




Depending on the geometry of the projection of the nodal lines onto the surface
BZ, the finite topological charge can lead to flat surface
bands~\cite{Burkov2011, Matsuura2013}. In Fig.~\ref{F1}(b), we plot the
spectrum as a function of $(k_y, k_z)$ for bands on the $(100)$ surface. The
flat bands are bounded by the nodal lines projected onto the surface BZ.
However, flat bands appear for all linking numbers, including the case $n = 0$
where the surface states are simply bounded by the two straight nodal lines.
Searching for ``pinch points'' on the surface bands is not sufficient to
confirm linked lines, as one can construct unlinked loops with non-planar
geometries where pinch points still appear. In the following, we suggest two
approaches to probe the linking number of nodal lines. First, we show the
Chern-Simons 3-form evaluated over the Brillouin zone is related to the linking
number. The Chern-Simons 3-form is related to the axion Lagrangian, a quantity
also present in Weyl semimetals. Future work may uncover how linked nodal
rings affect the magnetoelectric transport. Second, we compute the Landau level
spectrum and show the consequences of linked vs. unlinked nodal lines.

\section{Linking numbers and Chern-Simons theory}

We are interested in defining a topological invariant that distinguishes
whether the rings are linked or not. In the standard effective field theory of
3D topological insulators, the term of interest is the axion Lagrangian
Eq.~(\ref{Eq:axion}) with $\theta$ constrained by time-reversal symmetry to be
either zero or $\pi$. The $\theta$ angle can be computed from the Berry
connection via the Chern-Simons 3-form,
\begin{equation}
  \theta = \frac{1}{4\pi} \int_\text{BZ} {\rm Tr} (a \wedge da - \frac{2 i}{3} a \wedge a \wedge a).
\end{equation}
In the case of an abelian Chern-Simons theory, Polyakov and Witten showed that
this form is deeply related to the linking number $N$ of Wilson
loops~\cite{Witten1989}. The linking number is one of the most basic invariants
characterizing loops, and is computed using the Gauss linking integral,
\begin{equation}
  N(\L_\alpha, \L_\beta) = \frac{1}{4 \pi}
  \oint_{\L_\alpha }d x^i
  \oint_{\L_\beta} d y^j \epsilon_{ijk} \frac{(x-y)^k}{|x-y|^3},
\end{equation}
which determines the (signed) number of times two loops $\L_\alpha$
and $\L_\beta$ intertwine with one another. This theory has been
applied to solid state in the context of line-node superconductors, with the
line-nodes playing the role of Wilson loops~\cite{Lian2017}.
It is shown that
the linking number is related to the Chern-Simons form by
\begin{align}
  \frac{1}{4\pi} \int a \wedge d a = \pi \sum_{\alpha,\beta} \nu_\alpha \nu_\beta N(\L_\alpha, \L_\beta),
  \label{Eq:links}
\end{align}
where $\nu_\alpha$ and $\nu_\beta$ are the vorticities associated with the
nodal rings. We assume the self-linking number $N(\L_\alpha,\L_\alpha)$ is
zero. A heuristic sketch of the above equality is shown in the Supplemental 
Material \cite{SM}.

The relationship of the Chern-Simons form to the linking number is rather
subtle due to the nodal lines. Consider first the magnetic field $da$
associated with a nodal line $\L_\alpha$. Applying Gauss' theorem to
Eq.~(\ref{Eq:vorticity}) gives
\begin{equation}
  2\pi \nu_\alpha = \oint_{\C_\alpha} a =\int_{\S_\alpha} da,
\end{equation}
where $\S_\alpha$ is the oriented surface bounded by an arbitrary loop
$\C_\alpha$ encircling the nodal line $\L_\alpha$. This implies the magnetic
field must vanish everywhere except right along the nodal line,
\begin{equation}
  da = \sum_{\alpha}2 \pi \nu_\alpha
  \delta^{(2)} (\k_\perp- \L_\alpha) d k^1_\perp \wedge d k^2_\perp,
\end{equation}
where $\delta^{(2)} (\k_\perp - \L_\alpha)$ is the two-dimensional delta
function at the nodal line and $\k_\perp$ is the two component vector lying
perpendicular to the direction of $\L_\alpha$. Note the sign of the charge
$\nu_\alpha$ depends on the local choice of the coordinate system $\k_\perp$.
Since the magnetic field is nonzero only along the line node, the integral of
the Chern-Simons form over the BZ becomes a line integral of $\L_\alpha$:
\begin{align}
  \frac{1}{4\pi} \int_\text{BZ} a \wedge d a &= \frac{1}{2} \sum_{\alpha} \nu_\alpha \oint_{\L_\alpha} a \\
  & =\pi \sum_{\alpha,\beta} \nu_\alpha \nu_\beta N(\L_\alpha, \L_\beta).
\end{align}
Strictly, the integral $\oint_{\L_\alpha} a$, which encircles the ring
$\L_\beta$, is undefined since the Berry connection is ill-defined along
$\L_\alpha$. However, if one allows the integration path to be slightly
deformed away from the nodal line via $\L_\alpha \rightarrow \C_\beta$, the
integral becomes the topological charge of $\L_\beta$ multipled by the number
of times $\L_\alpha$ encircles the nodal line $\L_\beta$. For the model
described by Eq.~(\ref{Eq:f}), we have $\nu_\alpha \nu_\beta= -1/4$, leading to
$\theta = N\pi/2$, where $N$ is the linking number between the two nodal rings.
 This quantization condition is a result of chiral symmetry: in Weyl-link
  semimetals, the Chern-Simons 3-form is reduced to the Chern-Simons 1-form
  because the two-dimensional delta functions along the nodal lines reduces the
  integral over the Brillouin zone to line integrals. In the presence of chiral
  symmetry, the Chern-Simons 1-form is quantized~\cite{ryuNJP10}. Combined with
  the $\pm 1/2$ topological charge of the nodal lines, Chern-Simons 3-form is
  quantized as $\theta = N\pi/2$.


In two-band models, the non-vanishing linking number derived from $\int a da$
arises from the divergent behavior of the Berry connection $a$ at the nodal
lines. In the presence of finite mass term $m\sigma_z$, the Berry connection
and Berry curvature are well-defined in the entire BZ,
\begin{align}
  a  &= \frac{g\partial_{k_i}f - f\partial_{k_i}g}{2 \xi(\xi + m)} dk_i, \\
  da &= -\frac{m}{\xi^3} (\partial_{k_j} f \partial_{k_k} g) d k_j \wedge d k_k,
\end{align}
where $\xi=\sqrt{f^2+g^2+m^2}$. One can directly compute $\int ada$
unambiguously and the integral uniformly vanishes since $a$ and $da$ are
perpendicular to each other. Mathematically, it can be shown that for a
well-defined Berry connection $a$, at least four bands are necessary for a
finite value of $\int ada$.

A model with two linked Weyl rings need not have a nonzero Chern-Simons 3-form
because the two rings may originate from decoupled blocks of the hamiltonian.
Specifically, {we consider a minimal four-band model}
\begin{equation}
  \H(\k) = \begin{pmatrix}
    \H_1 (\k) & \\
    & \H_2 (\k)
  \end{pmatrix},
\end{equation}
where $\H_1 (\k)$ produces the set of rings $\{\L_\alpha^1\}$ and $\H_2 (\k)$
similarily produces $\{ \L_\alpha^2 \}$. The Berry connection over the filled
bands $a_{\alpha\beta}$ is a diagonal $2\times 2$ matrix with elements
$a_{\alpha \alpha}=\langle u_\alpha^- | \partial u_\alpha^- \rangle d k_i$ on
the diagonal. Here, $|u_\alpha^- \rangle$ is the occupied band for
$\H_{\alpha}(\k)$, $\alpha = 1, 2$. The Chern-Simons form then decomposes into
a simple sum
\begin{align}
  \int {\rm Tr} (a da) = \int a_{11} da_{11} +  \int a_{22} da_{22}.
\end{align}
Even though the two sets of rings $\{\L_\alpha^1\}$ and $\{\L_\beta^1\}$ may be
linked in the BZ, the links across the two sets do not contribute to the
Chern-Simons 3-form. Thus, there can be trivial WLSMs with $\theta = 0$.

In 3D topological insulators, the $\theta$-term in the axion Lagrangian plays
an essential role in their response to electromagnetic fields. In gapped
phases, the derivation of the axion Lagrangian is a well-defined procedure:
integrate out the fermions to leave a low energy theory in terms of the
physical electromagnetic gauge field. In gapless phases, fermions coexist with
the gauge field at the nodal points and lines, so their response may not be
separable from the topological effects of the axion Lagrangian. Nevertheless,
the $\theta$-term has been computed in Weyl semimetals via integration of the
fermions~\cite{Chen2013}. We propose that a similar procedure may be fruitful
for WLSM and may relate the linking number of the nodal lines to transport.

\section{Landau level spectrum}
The Landau level spectrum provides an independent probe of the linking of nodal
rings. Consider the continuous version of the model in Eq.~(\ref{Ham}),
derived by expanding the functions $f$ and $g$ up to second order in
momentum:
\begin{align}
  f(\k) &\to 3 - \frac{1}{2}(k_x^2+k_y^2) \\
  g(\k) &\to \left(1-\frac{1}{2}(n k_z)^2 +n k_z\right) k_x \notag \\
  & \qquad + \left(1-\frac{1}{2}(nk_z)^2 - nk_z\right) k_y.
\end{align}
For the specific case when the magnetic field lies along the $z$-direction, we
do not expand the $k_z$ terms, keeping them as $\cos nk_z$ and $\sin n k_z$.
In the presence of the magnetic field $B_\gamma$ along $\gamma$-direction
($\gamma=x,y,z$), the momentum $k_\gamma$ is a good quantum number and the
Hamiltonian can be parametrized as $\mathcal{H}(k_\gamma,B_\gamma)$. By
introducing the ladder operators $a$ and $a^\dagger$, we can write the
conjugate momenta as $\Pi_\alpha=k_\alpha-eA_\alpha=\frac{1}{\sqrt{2} l_B}
(a^\dagger + a)$ and $\Pi_\beta=k_\beta-eA_\beta=\frac{1}{\sqrt{2} i l_B}
(a^\dagger - a)$ with $l_B=1/\sqrt{eB_\gamma}$ being the magnetic length. Here,
$(\alpha,\beta,\gamma)$ form a cyclic permutation of $(x, y, z)$. The
Hamiltonian is a two by two matrix with parameters $(k_\gamma, B_\gamma)$ and
ladder operators $(a, a^\dagger)$. The eigenfunction satisfies
$\mathcal{H}(k_\gamma, B_\gamma)\Psi= E(k_\gamma, B_\gamma)\Psi$, with $\Psi
=(\sum_m \alpha_m | m\rangle, \sum_m \beta_m | m\rangle)^{\rm T}$, where
$|m\rangle$ is the state that satisfies $a^\dagger |m\rangle = \sqrt{m+1}
|m+1\rangle $ and $a |m\rangle = \sqrt{m} |m-1 \rangle$.

\begin{widetext}
The Landau level spectrum can be obtained by solving coupled equations of the
form
\begin{align}
  \mathbb{A}_m \beta_{m+3}+\mathbb{B}_m \beta_{m+2}+\mathbb{C}_m \beta_{m+1}+\mathbb{D}_m \beta_{m}+\mathbb{E}_m \beta_{m-1}+\mathbb{F}_m \beta_{m-2}+\mathbb{G}_m \beta_{m-3} &= E(k_\gamma, B_\gamma) \alpha_m, \notag\\
  \mathbb{G}^*_{m+3} \alpha_{m+3}+\mathbb{F}^*_{m+2} \alpha_{m+2}+\mathbb{E}^*_{m+1} \alpha_{m+1}+\mathbb{D}^*_m \alpha_{m}+\mathbb{C}^*_{m-1} \alpha_{m-1}+\mathbb{B}^*_{m-2} \alpha_{m-2}+\mathbb{A}^*_{m-3} \alpha_{m-3} &= E(k_\gamma, B_\gamma) \beta_m,
  \label{Eq: coupled}
\end{align}
where $m \geqslant 0$ and $\mathbb{A}_n$, $\mathbb{B}_n$, $\mathbb{C}_n$,
$\mathbb{D}_n$, $\mathbb{E}_n$, $\mathbb{F}_n$, and $\mathbb{G}_n$ are the
coefficients depend on the direction of the magnetic field (see Supplemental 
  Material \cite{SM}).
\end{widetext}

\begin{figure}
  \includegraphics[width=\columnwidth] {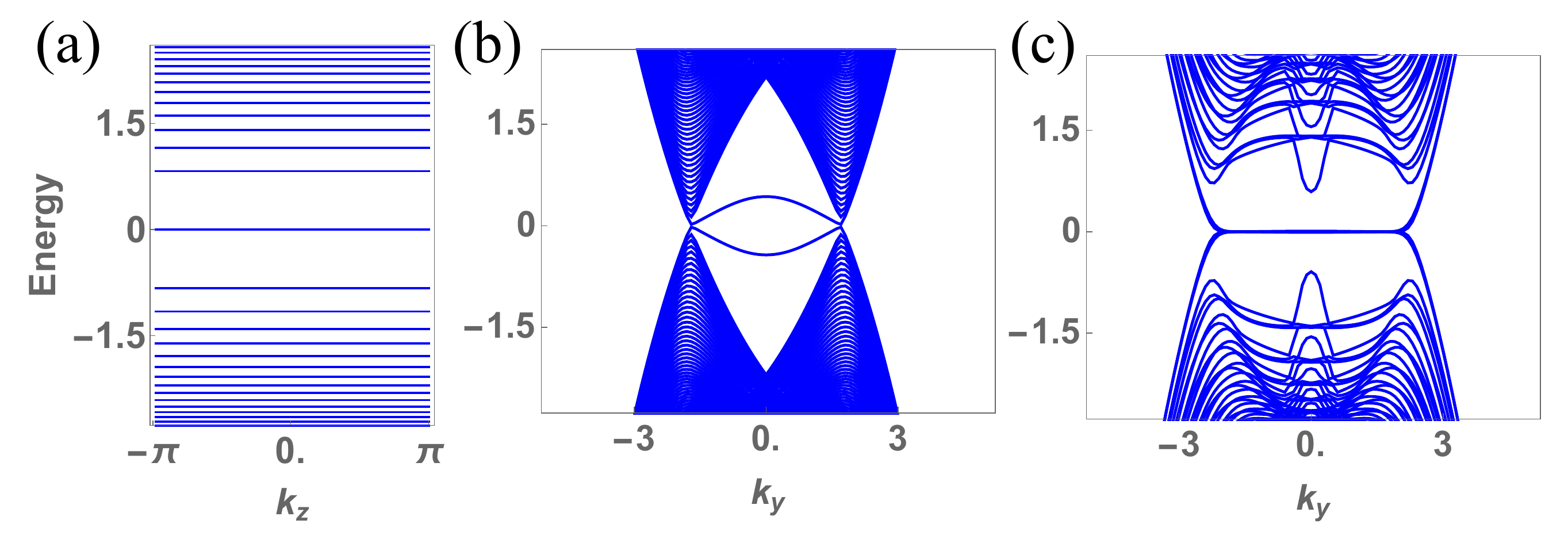}
  \caption{The Landau level spectra as a function of momentum. (a) The magnetic
    field along $z$-direction with arbitrary linking parameter $n$, (b)the
    magnetic field along $y$-direction with linking parameter $n=0$ and (c) the
    magnetic field along $y$-direction with linking parameter $n=1$.}
  \label{Landau}
\end{figure}

For a magnetic field oriented along the $z$-direction, the Landau level
spectrum is dispersionless as a function of $k_z$ and independent of the
linking parameter $n$. As shown in Fig.~\ref{Landau}(a), the level spacing is
proportional to $\sqrt{m}$, where $m$ is the Landau level index. At each fixed
$k_z$ slice, the system is effectively a two-dimensional spinless graphene
sheet with two Weyl points which give rise to the doubly degenerate Landau
level spectrum $E_m = {\rm sgn}(m) e B_z \sqrt{|m|} $. Since the distance
between the two Weyl points in the $k_x$-$k_y$ plane is the same for any
linking parameter $n$, the Landau level spectrum is independent of the linking
parameter $n$.

For a magnetic field oriented along the $y$-direction, the Landau level
spectrum strongly depends on the linking parameter $n$. For unlinked lines
($n=0$), the Landau level spectrum $E_m(k_y)$ closes at the nodal lines
locations projected on $k_y$ axis. In between these two gap closing points,
there are two dispersive mid-gap states, as shown in Fig.~\ref{Landau}(b). For
linked lines ($n=1$), the Laudau level spectrum has four dispersionless
zero-energy modes in between the two gap closing points [Fig.~\ref{Landau}(c)].
This four-fold degeneracy originates from the four Weyl points that pierce each
fixed $k_y$ slice that lies between the two gap closing points, and each Weyl
point contributes one zero mode. {These zero modes 
are stabilized by chiral symmetry.} By changing the orientation of the magnetic
field, the location of the nodal lines can be mapped out and their linking
number determined, similar to the use of quantum oscillation measurements for
mapping out Fermi surface geometry.

\section{Conclusion and outlook}

We introduce a two-band model that generates a family of WLSMs with arbitrary
linking number. We show that the linking number of the nodal rings is related
to the Chern-Simons 3-form. One possible materials candidate is a 3D carbon
allotrope composed of pentagonal rings proposed by Zhong, {\it
  et.al,}~\cite{YPChen_2017}. The electronic structure of the 3D carbon
allotrope under strain exhibits a Weyl-link semimetallic phase. We note that
the minimal model for the 3D carbon allotrope is a three-band model, where the
two nodal rings originate from the crossing of two different pairs of bands. In
the future, we hope to apply this construction based on the intersection of two
hypersurfaces to produce semimetals with trefoil knots~\cite{PYCCHY}.

\section{Note  added}
During the preparation of this manuscript, we became aware of related work by
Chen, {\it et.al.}~\cite{Chen2017} and Yan, {\it et. al.}~\cite{Yan2017}.

\section{Acknowledgements}
The authors would like to thank Yuanping Chen, Elio K\"onig, Piers Coleman and
David Vanderbilt for valuable discussions. P.-Y. C. was supported by the
Rutgers Center for Materials Theory postdoctoral grant. C.H.Y. was supported in
part by the Center for Emergent Superconductivity, an Energy Frontier Research
Center funded by the US Department of Energy, Office of Science, Office of
Basic Energy Sciences under Award No. DEAC0298CH1088.

\bibliography{bibliography}

\clearpage

\pagebreak

\newpage

\begin{center}

\noindent{\Large Supplementary Material for ``Weyl-link semimetals". }
\end{center}

\subsection{Topological invariants for linking number}
\begin{figure}[htbp]
  \centering
  \includegraphics[height=5 cm] {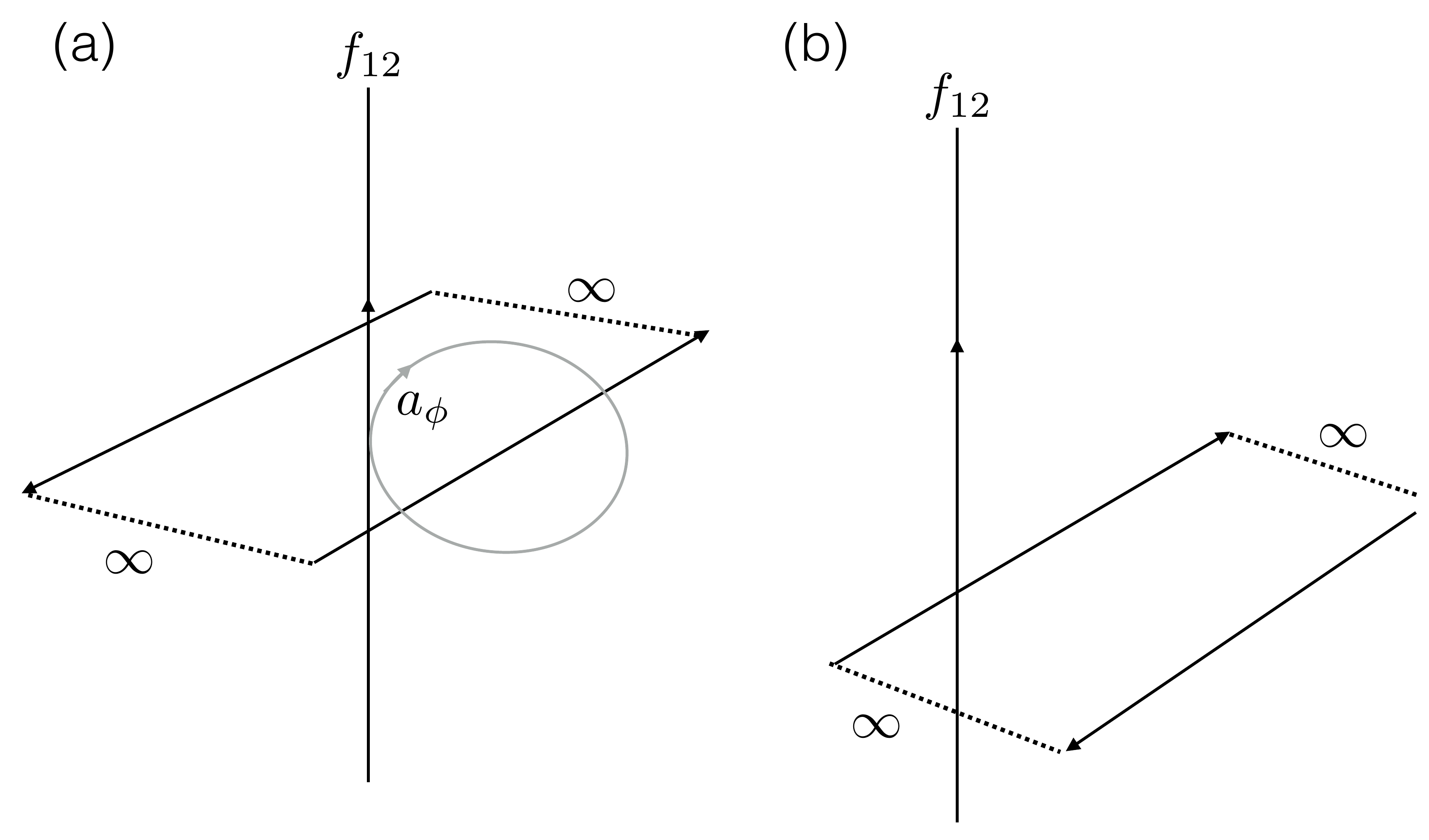}
  \caption{(a) Linked configuration: one nodal line along $k_z$ direction and
    two other nodal lines on the $k_x-k_y$ plane where they connected at
    infinity indicated by dashed lines. The vertical nodal line is encircled by
    the horizontal nodal lines. (b) Unlinked configuration: the vertical nodal
    line is not encircled by the horizontal nodal lines.}
  \label{Fig_2}
\end{figure}

We demonstrate the relation between the linking number and the Chern-Simons 3-form
in Eq. (9) by considering a simple geometry shown in Fig. \ref{Fig_2}.
Let us start with a Hamiltonian, $\H=k_1 \sigma_1 + k_2\sigma_2 +m \sigma_3$.
When the mass term vanishes, the model exhibits a nodal line along $k_3$ direction.
The occupied band is
\begin{align}
  |u^-(\k) \rangle = \frac{1}{\sqrt{\xi (\xi +m)}} \left(\begin{array}{c}-k_1+i k_2 \\\xi + m\end{array}\right),
\end{align}
where $\xi = \sqrt{k_1^2 +k_2^2 +m^2}$. Assuming $m>0$, the Berry connection
\begin{align}
  a_1=\frac{k_2}{2 \xi (\xi+m)},\quad a_2 = \frac{k_1}{2 \xi(\xi+m)},
\end{align}
and the Berry curvature
\begin{align}
  f_{12} = \partial_{k_1} a_2 - \partial_{k_2} a_1 = \frac{m}{2 \xi^3}.
\end{align}

When we take the limit $m \to 0$, the Berry curvature
\begin{align}
  \lim_{m \to 0}f_{12} = -\pi \delta^{(2)} (k_1,k_2),
\end{align}
and the Berry connection in the limit $m \to 0$ has only the $\phi$ component in the cylindrical coordinate
\begin{align}
  a_{\phi} =& -\frac{1}{2 k}, \quad k \neq 0
\end{align}
where $k=\sqrt{k_1^2+k_2^2}$.

The topological charge of this nodal line is $\nu = \frac{1}{2 \pi i} \oint
a_\phi k d \phi = \frac{1}{2}$. Now we consider two perpendicular nodal lines
separated with distance $k_0$ as shown in Fig. \ref{Fig_2}. In Fig.
\ref{Fig_2}(a), one nodal line is encircling the other one, where the dashed
lines are at infinite. On the other hand, in Fig. \ref{Fig_2}(b), these two
lines are not linked. We can separate the Berry connection into two
$a=a_1+a_2$, where $a_i$ is the Berry connection generated by $i$-th nodal
line. We have
\begin{align}
  \int a da =  \int  (a_1+a_2) (da_1+da_2)= \int  a_1 d a_2 + \int  a_2 d a_1.
\end{align}
Here $\int a_i da_i = 0 $ in the presence of infinitesimal mass term. Let us
suppose $a_1 = -\frac{1}{2 k}$ is generated by the nodal line along $y$ axis
and $d a_2 = f_2 = -\pi \delta(k_1) \delta(k_2+k_0)$ is generated by the nodal
line along $z$ direction at $(k_x,k_y) = (0,-k_0)$.
\begin{align}
  \int a_1 d a _2 =\frac{\pi}{2 k_0 } \int d{k_3} \frac{k_0^2}{k_0^2+k_3^2} =\frac{\pi^2}{2}.
\end{align}

For the configuration shown in Fig. \ref{Fig_2}(a), $\int a da = 4
*\frac{\pi^2}{2}=2 \pi^2$. From Eq. (9), the Gauss linking number is
\begin{align}
  N(\L_b, \L_c) = \frac{1}{2\pi \nu_b*\nu_c}\frac{1}{4 \pi}\int a da =1,
\end{align}
where $\nu_b=\nu_c=1/2$.

On the other hand, for the configuration shown in Fig. \ref{Fig_2}(b), $\int a
da = 2 *\frac{\pi^2}{2}- 2 *\frac{\pi^2}{2} = 0$, which indicates the Gauss
linking number is zero.

\subsection{Landau level spectrum}
We consider two orientations of the magnetic field: (a) along $z$-direction and (b) along $y$-direction.
The Hamiltonian in the presence of magnetic field $B_\gamma$ ($\gamma=z, y$) can be parametrized as
$\mathcal{H}(k_\gamma,B_\gamma)$.  We can introduce 
the ladder operators $a$ and $a^\dagger$ such that
 the conjugate momenta can be written as $\Pi_\alpha=k_\alpha-eA_\alpha=\frac{1}{\sqrt{2} l_B} (a^\dagger + a)$ and $\Pi_\beta=k_\beta-eA_\beta=\frac{1}{\sqrt{2} i l_B} (a^\dagger - a)$
with $l_B=1/\sqrt{eB_\gamma}$ being the magnetic length. Here $\alpha$, $\beta$ and $\gamma$ form a cyclic order in $[x, y, z]$. 
We solve the Landau level spectrum by introducing a trial wavefunction $\Psi =(\sum_m \alpha_m | m\rangle, \sum_m \beta_m | m\rangle)^{\rm T}$
with $| m\rangle$ being the state that satisfies $a^\dagger |m\rangle = \sqrt{m+1} |m+1 \rangle $ and $a |m\rangle = \sqrt{m} |m-1 \rangle$.
The eigenfunction satisfies $\mathcal{H}(k_\gamma, B_\gamma)\Psi= E(k_\gamma, B_\gamma)\Psi$.
In the case that the magnetic field pointing along $z$-direction, the Hamiltonian (we expend $\sin k_{x/y} \to k_{x/y}$ and $\cos k_{x/y} \to 1-\frac{1}{2} k^2_{x/y}$) is
\begin{widetext}
\begin{align}
\mathcal{H}(k_z, B_z) =&  \left[ 3-\frac{eB_z}{2}(2a^\dagger a +1) \right] \mathbb{I}_{2 \times 2}     \notag\\
&+\sqrt{e B_z} \left(\begin{array}{cc}0 & -i[e^{i (n k_z -\frac{\pi}{4})}a^\dagger+
e^{-i (n k_z -\frac{\pi}{4})} a]  \\ i [e^{-i (n k_z -\frac{\pi}{4})}a+
e^{i (n k_z -\frac{\pi}{4})} a^\dagger] & 0\end{array}\right).
\end{align}

We solve the spectrum by expressing the trial wavefunction $\Psi =(\sum_m \alpha_m | m\rangle, \sum_m \beta_m | m\rangle)^{\rm T}$
with $| m\rangle$ being the state that satisfies $a^\dagger |m\rangle = \sqrt{m+1} |m+1 \rangle $ and $a |m\rangle = \sqrt{m} |m-1 \rangle$.
We have the following coupled equations
\begin{align}
 &[3-\frac{eB_z}{2}(2m +1)] \beta_m-i\sqrt{e B_z}[e^{i (n k_z - \frac{\pi}{4})}\sqrt{m} \beta_{m-1}+ e^{-i (n k_z - \frac{\pi}{4})}\sqrt{m+1} \beta_{m+1}] = E(k_z, B_z) \alpha_m, \notag\\
 &[3-\frac{eB_z}{2}(2m +1)] \alpha_m+i\sqrt{e B_z}[e^{i (n k_z - \frac{\pi}{4})}\sqrt{m} \alpha_{m-1}+e^{-i (n k_z - \frac{\pi}{4})}\sqrt{m+1} \alpha_{m+1}] = E(k_z, B_z) \beta_m.
\end{align}
The coefficients in Eq. (21) in the main text, $\mathbb{A}_{m}$, $\mathbb{B}_{m}$, $\mathbb{F}_{m}$, and $\mathbb{G}_{m}$ are vanishing
and $\mathbb{C}_{m}= - i \sqrt{e B_z} e ^{-i (n k_z - \frac{\pi}{4})}$,  $\mathbb{D}_{m}= 3 - \frac{e B_z}{2} (2 m+1)$,  $\mathbb{D}_{m}= - i \sqrt{e B_z} e ^{i (n k_z - \frac{\pi}{4})}$.
We observe the spectrum $E(k_z, B_z)$ is completely flat as a function of $k_z$ and is independent of $n$.
These flat bands come from connecting the conventional Landau level in two dimensional Weyl cone on the $k_z$ slices.
Since the Landau level in two dimensional Weyl cones are not sensitive to the location of the cones, the spectrum $E(k_z, B_z)$ is independent on the linking parameter $n$.

In the case that the magnetic field is along $y$-direction, the Hamiltonian is
\begin{align}
\mathcal{H}(k_y, B_y) =&  \left[ 3-\frac{1}{2} k_y^2-\frac{eB_y}{4}(2a^\dagger a +1) \right] \mathbb{I}_{2 \times 2}   
+k_y\left[1- \frac{n^2 e B_y}{4}  (2a^\dagger a +1)\right] \sigma_y  \notag\\
&+ \frac{e B_y}{4}(1-2n) \left(\begin{array}{cc}0 & a^\dagger a^\dagger \\a a & 0\end{array}\right)
+\frac{n^2 k_y e B_y}{4}\left(\begin{array}{cc}0 & i a^\dagger a^\dagger \\-i a a & 0\end{array}\right)   \notag\\
&+\sqrt{\frac{e B_y}{{2}}} \left[ (-1+\frac{n^2 e B_y}{4}) \left(\begin{array}{cc}0 & a^\dagger \\  a & 0\end{array}\right)
+n k_y\left(\begin{array}{cc}0 & i a^\dagger \\  -i a & 0\end{array}\right) + \frac{n^2 e B_y}{4}  \left(\begin{array}{cc}0 & a^\dagger a^\dagger a \\a^\dagger a a & 0\end{array}\right)  \right]  \notag\\
&+\sqrt{\frac{e B_y}{{2}}} \left[ (1-\frac{n^2 e B_y}{4}) \left(\begin{array}{cc}0 & a\\  a^\dagger  & 0\end{array}\right)
+n k_y\left(\begin{array}{cc}0 & i a \\  -i a^\dagger  & 0\end{array}\right) - \frac{n^2 e B_y}{4}  \left(\begin{array}{cc}0 & a^\dagger a a \\a^\dagger a^\dagger  a & 0\end{array}\right)  \right]   \notag\\
&+\sqrt{\frac{e B_y}{{2}}} \frac{n^2 e B_y}{4} \left(\begin{array}{cc}0 & a^\dagger a^\dagger a^\dagger - aaa \\ a a a - a^\dagger a^\dagger a^\dagger& 0\end{array}\right).
\end{align}

Here we symmetrize $\Pi_z^2\Pi_x \to \frac{1}{2} [\Pi_z^2 \Pi_x +\Pi_x \Pi_z^2 ]$.
The coefficients from coupled equations  in Eq. (21) in the main text from the trial wavefunction $\Psi =(\sum_m \alpha_m | m\rangle, \sum_m \beta_m | m\rangle)^{\rm T}$
with $| m\rangle$ being the state that satisfies $a^\dagger |m\rangle = \sqrt{m+1} |m+1 \rangle $ and $a |m\rangle = \sqrt{m} |m-1 \rangle$ are
\begin{align}
&\mathbb{A}_m=-\sqrt{\frac{e B_y}{{2}}} \frac{n^2 e B_y}{4} \sqrt{(m+1)(m+2)(m+3)},   \notag\\
&\mathbb{B}_m=\frac{ e B_y}{4} (1+2n+i n^2 k_y) \sqrt{(m+1)(m+2)},\notag\\
&\mathbb{C}_m= \sqrt{\frac{e B_y}{{2}}} \left[ 1+ i n k_y - \frac{n^2 e B_y }{4}(1+m)      \right]    \sqrt{(m+1)},\notag\\
&\mathbb{D}_m= 3 - \frac{1}{2}k_y^2 -\frac{e B_y}{4} (2m+1)- i k_y\left[1- \frac{n^2 e B_y}{4} (2m+1)\right], \notag\\
&\mathbb{E}_m= \sqrt{\frac{e B_y}{{2}}} \left[ -1+ i n k_y + \frac{n^2 e B_y }{4}m     \right]    \sqrt{m},\notag\\
&\mathbb{F}_m=\frac{ e B_y}{4} (1-2n+i n^2 k_y) \sqrt{m(m-1)},\notag\\
&\mathbb{G}_m=\sqrt{\frac{e B_y}{{2}}} \frac{n^2 e B_y}{4} \sqrt{m(m-1)(m-2)}.
\end{align}

In the case that the linking parameter $n=0$, the Landau level spectrum $E_m(k_y)$ has two gap closing points 
where two nodal lines are located at. On the other hand, in the case that the linking parameter $n=1$, there are four zero-energy 
modes in the Landau spectrum $E_m(k_y)$. The existence of zero-energy modes are originated from the zero-energy Landau level of Weyl points at a fixed $k_y$ slide.

\end{widetext}

\end{document}